\documentstyle[12pt, doublespace]{article}
\newcommand{\journal}[4]{{\em #1~}#2\,(19#3)\,#4;}

\newcommand{\ijmp}{\journal {Int. J. Mod. Phys.}}

\newcommand{\pr}{\journal {Phys. Rev.}}

\newcommand{\cmp}{\journal {Comm. Math. Phys.}}
\newcommand{\cqg}{\journal {Class. Quantum Grav.}}

\newcommand{\np}{\journal {Nucl. Phys.}}
\newcommand{\pl}{\journal {Phys. Lett.}}

\newcommand{\prep}{\journal {Phys. Reports}}

\makeatletter
\@addtoreset{equation}{section}
\makeatother

\def\Lp{\displaystyle{\biggl(}}
\def\Rp{\displaystyle{\biggr)}}
\def\LP{\displaystyle{\Biggl(}}
\def\RP{\displaystyle{\Biggr)}}

\newcommand{\lc}{\left[}\newcommand{\rc}{\right]}

\newcommand{\D}{\Delta}
\renewcommand{\a}{\alpha}
\renewcommand{\b}{\beta}
\renewcommand{\d}{\delta}
\newcommand{\e}{\varepsilon}
\newcommand{\eb}{\bar\varepsilon}
\newcommand{\f}{\phi}
\newcommand{\F}{\Phi}
\renewcommand{\P}{\Psi}
\newcommand{\Pb}{\bar\Psi}

\newcommand{\g}{\gamma}

\newcommand{\x}{\xi}
\renewcommand{\l}{\lambda}
\newcommand{\lb}{\bar\lambda}
\newcommand{\Lb}{\bar\Lambda}
\renewcommand{\L}{\Lambda}
\newcommand{\m}{\mu}
\newcommand{\n}{\nu}

\renewcommand{\o}{\omega} \renewcommand{\O}{\Omega}
\newcommand{\p}{\psi}
\renewcommand{\pb}{\bar\psi}

\newcommand{\s}{\sigma} \renewcommand{\S}{\Sigma}
\newcommand{\Sh}{\hat\Sigma}
\renewcommand{\t}{\theta}

\newcommand{\BB}{{\cal B}}
\newcommand{\BSh}{\hat{\cal B}_\Sigma}
\newcommand{\BS}{{\cal B}_\Sigma}

\newcommand{\FF}{{\cal F}}

\newcommand{\HH}{{\cal H}}

\newcommand{\PP}{{\cal P}}
\newcommand{\QQ}{{\cal Q}}

\newcommand{\SSh}{\hat{\cal S}}
\newcommand{\SS}{{\cal S}}

\newcommand{\complex}{{\kern .1em {\raise .47ex
\hbox {$\scriptscriptstyle |$}}
    \kern -.4em {\rm C}}}
\newcommand{\real}{{{\rm I} \kern -.19em {\rm R}}}
\newcommand{\rational}{{\kern .1em {\raise .47ex
\hbox{$\scripscriptstyle |$}}
    \kern -.35em {\rm Q}}}
\renewcommand{\natural}{{\vrule height 1.6ex width
.05em depth 0ex \kern -.35em {\rm N}}}

\newcommand{\cb}{{\bar c}}
\newcommand{\half}{\frac 1 2}
\newcommand{\pa}{\partial}
\newcommand{\pad}[2]{{\frac{\partial #1}{\partial #2}}}
\newcommand{\fud}[2]  {{\displaystyle{\frac{\delta #1}{\delta #2}}}}

\newcommand{\ie}{{{\em i.e.}\ }}

\newcommand{\sla}{\raise.15ex\hbox{$/$}\kern -.57em}

\newcommand{\twiddle}{\lower.9ex\rlap{$\kern -.1em\scriptstyle\sim$}}

\renewcommand{\=}{&=&} 

\newcommand{\eq}{\begin{equation}}
\newcommand{\eqn}[1]{\label{#1}\end{equation}}
\newcommand{\eea}{\end{eqnarray}}
\newcommand{\eqa}{\begin{eqnarray}}
\newcommand{\eqan}[1]{\label{#1}\end{eqnarray}}
\newcommand{\ba}{\begin{array}}
\newcommand{\ea}{\end{array}}
\newcommand{\eqac}{\begin{equation}\begin{array}{rcl}}
\newcommand{\eqacn}[1]{\end{array}\label{#1}\end{equation}}

\renewcommand{\pad}[2]{{\displaystyle{\frac{\partial #1}{\partial #2}}}}

\newcommand{\intx}{\int d^4 \! x \, }

\begin{document}
\def\ftoday{{\sl  \number\day \space\ifcase\month
\or Janvier\or F\'evrier\or Mars\or avril\or Mai
\or Juin\or Juillet\or Ao\^ut\or Septembre\or Octobre
\or Novembre \or D\'ecembre\fi
\space  \number\year}}



\titlepage
{\singlespace

\begin{center}

{\huge  Off--shell formulation of N = 2 Super Yang--Mills theories
         coupled to matter without auxiliary fields         }

\vspace{1cm}

{\Large Nicola Maggiore}\footnote{Supported in part
by the Swiss National Science Foundation.}

{\it D\'epartement de Physique Th\'eorique --
     Universit\'e de Gen\`eve\\24, quai E. Ansermet -- CH-1211 Gen\`eve
     4\\Switzerland}

\end{center}

\vspace{2.5cm}

\begin{center}
\bf ABSTRACT
\end{center}

{\it
N=2 supersymmetric Yang--Mills theories coupled to matter are
considered in the Wess--Zumino gauge. The supersymmetries are realized
nonlinearly and the anticommutator between two susy charges gives, in
addition to
translations, gauge transformations and equations of motion. The difficulties
hidden in such an algebraic structure are well known: almost always auxiliary
fields can be introduced in order to put the formalism off--shell, but still
the field dependent gauge transformations give rise to an infinite dimensional
algebra quite hard to deal with. However, it is possible to avoid all these
problems by collecting into an unique nilpotent operator all the symmetries
defining the theory, namely ordinary BRS, supersymmetries and translations.
According to this method the role of the auxiliary fields is covered by the
external sources coupled, as usual, to the nonlinear variations of the quantum
fields. The analysis is then formally reduced to that of ordinary Yang--Mills
theory.}

\vfill
\noindent
hep-th/9412092\\
UGVA-DPT-94-12-870 \hfill December 1994      }
\newpage

\section{Introduction}

Amongst the Supersymmetric Yang -- Mills (SYM) theories, the extended
$N=2$ models~\cite{old} play a particular role, because of the number
of exact results that can be extracted from them. For instance,
a new mechanism for chiral symmetry breaking deriving from the condensation
of magnetic monopoles has been investigated for $N=2$ SYM in~\cite{wit}, where
also an analysis of the electric magnetic duality is done.
Another reason of interest is their
generality: in fact, the reduction to the $N=1$ case being
straightforward, also the maximally extended $N=4$ SYM can be considered  as
a $N=2$ theory, with matter in the adjoint representation of the gauge
group.

In the Wess--Zumino gauge, the supersymmetry transformations are
nonlinear, and consequently the major problem affecting
the model is related to its algebraic structure --~described
in Section 2~--, which involves equations of motion and field dependent
gauge transformations. This gives rise to an infinite dimensional
algebra, even if auxiliary fields can be introduced to put the
formalism off--shell~\cite{bm}.

Another difficulty consists in the definition of a gauge fixing
term, which, in the usual framework, is BRS invariant by construction
as it is BRS exact, but on the other hand it cannot be supersymmetric.

These two problems of SYMs, namely the existence of an infinite
dimensional algebraic structure and the non invariance of the gauge
fixing term, turn out to be somehow related to each other, because they
can both be solved at the same time.

In Section 3 we show indeed that it is possible to define a generalized
BRS operator which sums up the usual BRS invariance, the supersymmetry
and the invariance of the theory under translations.
The method we are going to exploit is quite general and it is particularly
powerful in view of the renormalization of models exhibiting non trivial
algebraic structures, not only of the supersymmetric type. In that direction,
it has been already successfully
applied to topological~\cite{top},
supersymmetric~\cite{dix,whi}, ordinary~\cite{sch}
gauge field theories as well as to non gauge field theories~\cite{bbc}.
In particular in~\cite{whi}, this
technique has been used to study the renormalization
of the $N=1$ and $N=4$ case. Here we generalize the description to the $N=2$
model, as a preliminary and necessary step for the discussion of its quantum
extension~\cite{wp}.

The essence consists simply in collecting
all the symmetries of the theory into an unique operator.
Opportune transformations of the
ghost fields make this generalized BRS operator on--shell nilpotent,
allowing us to construct an invariant gauge--fixing term.
This permits us to achieve the main purpose of this paper, \ie to
completely determine the $N=2$ SYMs at the classical level, by means of
a classical gauge fixed action invariant
under a Slavnov--like operator which summarizes all the symmetries of the
theory.

In Section 4 we solve classically the constraints
which define the theory. At the same time we give the expressions of the
operators which will be used in the algebraic renormalization of this
model to constrain the counterterms and the possible anomalies~\cite{wp}.

Results and perspectives are briefly discussed in the concluding Section 5.

\section{The model}

The massless irreducible representations of $N=2$ supersymmetry in four
dimensions are realized by means of the Super Yang -- Mills (SYM) multiplet
and the matter multiplet

\begin{center}
\begin{tabular}{lrrrr}
SYM & $A_\mu^a$ & $\l_{\a i}^a$ & $A^a$ & $B^a$
\\
matter & $A^{iA}$ & $A^\ast_{iA}$ & $\psi^\a_A$ & $\bar\psi^A_\a$  \\
\end{tabular}
\end{center}

Here $A_\mu^a$ is the gauge field and $\l_{\a i}^a$ is a doublet of Majorana
spinors, for which the following Majorana condition holds
\eq
\l_{\a i}^a = (i\gamma_5 C)_{\a\b}\epsilon_{ij}\bar\l^{\b j}\ ,
\eqn{21}
where $C$ is the charge conjugation matrix
$(C\g^\m C^{-1} = -{\g^\m}^T)$
and $\epsilon_{ij}$ is the two--dimensional Levi--Civita tensor.
The spinless fields $A^a$ and $B^a$ are a scalar and a pseudoscalar
respectively;  the matter fields consist in a doublet of
complex scalar fields $(A^{iA},\  A^\ast_{iA})$ and two Dirac spinors
$(\psi^\a_A,\ \bar\psi^A_\a)$.

The index $\m$ describes minkowskian spacetime, $\a$ is the spinorial index,
which runs from $1$ to $4$, and $i=1,2$ is the supersymmetry index. The
$N=2$ SYM multiplet belongs to the adjoint representation of a gauge group $G$,
while the matter multiplet can be in any representation, which we assume
to be real. The indices $a$ and $A$ run over the corresponding
Lie algebras. As it is well known, when also the matter multiplet belongs to
the adjoint representation of the gauge group, $N=4$ supersymmetry is
recovered.

In the Wess -- Zumino gauge, the supersymmetry is realized
nonlinearly~\cite{sw}
\eqa
\delta A^{a} &=& \eb^i\l^a_i \nonumber \\
\delta B^{a} &=& i\eb^i\g_5\l^a_i \nonumber \\
\delta A^{a}_\m &=& \eb^i\g_\m  \l^a_i  \\
\delta \l^a_{\a i}   &=& \frac{1}{2} F^a_{\m\n}(\s^{\m\n}\e_i)_\a
                  -(D_\m A)^a(\g^\m\e_i)_\a
\nonumber  \\ &&                +i(D_\m B)^a(\g^\m\g_5\e_i)_\a
                  +if^{abc}A^bB^c(\g_5\e_i)_\a \nonumber
\eqan{22}
\eqa
\delta A^{iA} &=& \eb^i\psi^A \nonumber \\
\delta \psi^A_\a &=& -2(D_\m A^i)^A (\g^\m\e_i)_\a
                   +2(T^a)^A_B A^{iB} A^a \e_{\a i} \\
&&                   -2i(T^a)^A_B A^{iB} B^a (\g_5\e_i)_\a\ , \nonumber
\eqan{23}
where $\s_{\m\n}\equiv\frac{1}{2}[\g_\m,\g_\n]$, and $(T^a)^A_B$ are
the generators of the gauge group $G$.

As usual, the nonlinearities are in the spinor transformation
laws, which depend on the field strenght
\eq
F^a_{\m\n}\equiv\pa_\m A^a_\n
-\pa_\n A^a_\m +f^{abc}A^b_\m A^c_\n
\eqn{24}
 and on the covariant derivatives
\eqa
(D_\m X)^a &\equiv& \pa_\mu X^a + f^{abc}A^b_\m X^c \nonumber\\
(D_\m Y)^A &\equiv& \pa_\mu Y^A + (T^a)^A_B A^a_\m Y^B \ .
\eqan{25}

The relation between the supersymmetry charges
$Q_{\a i}$ and the above supersymmetry operator $\delta$ is
\eq
\delta \equiv \eb^{\a i}Q_{\a i}\ ,
\eqn{26}
namely $\e_{\a i}$ is an infinitesimal fermionic anticommuting Majorana
parameter
\eq
\e_{\a i} = (i\gamma_5 C)_{\a\b}\epsilon_{ij}\eb^{\b j}\ .
\eqn{27}

The action
\eq
S=S_{SYM} + S_{matter} + S_{int} \ ,
\eqn{28}
where
\eqa
S_{SYM} = \frac{1}{g^2} \intx \LP &&\!\!\!
    -\frac{1}{4}F^{a\m\n}F^a_{\m\n} +\frac{1}{2}(D^\m A)^a(D_\m A)^a
    +\frac{1}{2}(D^\m B)^a(D_\m B)^a
\nonumber\\
&&\!\!\!
    +\half\lb^{ai}\g^\m(D_\m\l_i)^a
    +\half f^{abc}(\lb^{bi}\l^c_i)A^a
\\
&&\!\!\!
    -i\half f^{abc}(\lb^{bi}\g_5\l^c_i)B^a
    -\half f^{amn}A^mB^nf^{apq}A^pB^q \RP\ , \nonumber
\eqan{29}
\eq
S_{matter}=\intx\LP
(D^\m A^i)^A(D_\m A^\ast_i)_A
-\half\pb_A\g^\m(D_\m\p)^A\RP\ ,
\eqn{210}
\eqa
S_{int} = \intx \LP &&\!\!\!
-(T^a)^A_B(\pb_A\l^a_i)A^{iB}
+(T^a)^A_B(\lb^{ai}\p^B)A^\ast_{iA}
\nonumber \\
&&\!\!\!
+(T^a)^A_D(T^b)^D_BA^aA^bA^\ast_{iA}A^{iB}
+(T^a)^A_D(T^b)^D_BB^aB^bA^\ast_{iA}A^{iB}
\\ &&\!\!\!
-i\half (T^a)^A_B(\pb_A\g_5\p^B)B^a
-\half (T^a)^A_B(\pb_A\p^B)A^a\RP \ , \nonumber
\eqan{211}
and $g^2$ is the only coupling constant of the theory, is invariant  under
supersymmetry
\eq
\delta S =0\ .
\eqn{212}

The algebra formed by the supersymmetry transformations~(2.2)
and~(2.3)
exhibits two
obstructions to its closure on the translations~:
\eqa
[\d_1,\d_2]\F &=&\!\!\! 2 (\eb^i_1\g^\m\e_{2i})\pa_\m\F
\nonumber \\
&&\!\!\!
+2\d_{\rm gauge}(\o^a)\F
\\
&&\!\!\!
+\ \mbox{field equations}\ ,\nonumber
\eqan{213}
where $\F$ stands for all the fields of the theory.

The second term on the r.h.s. of equation~(2.13) is a gauge transformation
with a {\it field dependent} parameter~$\o^a$
\eq
\o^a\equiv
(\eb^i_1\g^\m\e_{2i})A_\m + i(\eb^i_1\g_5\e_{2i})B - (\eb^i_1\e_{2i})A\ ,
\eqn{214}
and the last term on the r.h.s of~(2.13) is a contact term present only
when $\F$ is one of the spinors of the theory, namely $\l$ or $\p$.

\section{The strategy}

The algebraic structure described by~(2.13) is typical of nonlinearly realized
supersymmetries and the general attitude when dealing with it is to put the
formalism off--shell firstly, by adding a
suitable number of auxiliary fields.

This causes a few drawbacks, the first of which, as it is apparent
from~(2.13),
is related to the fact that still
one is left with an algebra closing on translations
modulo the field dependent gauge transformations~(2.14).
This algebra is infinite--dimensional
and requires an infinite number of composite operators of increasing negative
dimensions in order to be controlled. As a consequence the discussion of the
renormalization of the model, which is our ultimate aim~\cite{wp}, becomes
difficult, if not impossible~\cite{bm}. In addition to that, no gauge fixing
term can be found
which is invariant under supersymmetry and therefore
one has also to deal with the
consequent breaking. Finally, not always auxiliary fields
can be introduced to eliminate the field equations from the algebra~(2.13).
In fact, if the matter multiplet belongs to the adjoint representation of the
gauge group, the theory can be interpreted as having a $N=4$ supersymmetry,
for which no off--shell formulation through
auxiliary fields is known~\cite{sw}.

These are the reasons why
we prefer to follow here an approach which does not rely on the
existence of auxiliary fields.
In order to do that, we first concentrate
on the elimination
from the algebra of the field dependent gauge transformations~(2.14),
temporarily
disregarding the presence of the equations of motion in~(2.13). We shall
get rid
of them subsequently, without introducing auxiliary fields.

Besides being supersymmetric, the action S~(2.8) is left invariant under
the usual BRS transformations
\eqa
s A^{a} &=& f^{abc} c^b A^c \nonumber\\
s B^{a} &=& f^{abc} c^b B^c  \nonumber\\
s A^{a}_\m &=& -(D_\m c)^a \nonumber\\
s \l^a_{\a i} &=& f^{abc} c^b \l_{\a i}^c \nonumber\\
s A^{iA} &=& (T^a)^A_B c^a A^{iB} \\
s \psi^A_\a &=& (T^a)^A_B c^a \p^B_\a \nonumber\\
s c^a &=& \half f^{abc}c^bc^c \nonumber\\
s \cb^a &=& b^a \ \ \  s b^a = 0 \ ,\nonumber
\eqan{31}
where $c$, $\cb$ and $b$ are respectively the ghost, the
antighost and the Lagrange
multiplier, which is introduced to implement the gauge condition
\eq
\pa^\m A^a_\m =0\ .
\eqn{32}

The BRS transformations~(3.1) are nilpotent
\eq
s^2=0\ ,
\eqn{33}
and they commute with the supersymmetry transformations~(2.2)-(2.3),
 trivially
extended to $c$, $\cb$ and $b$
\eq
\lc s,\d\rc =0\ .
\eqn{34}

The canonical dimensions and the Faddeev--Popov charges of the quantum fields
are reported in Table 1
\begin{center}
\begin{tabular}{|l|r|r|r|r|r|r|r|r|r|}\hline
 & $A_\mu^a$ & $\l_{\a i}^a$ & $A^a$ & $B^a$ & $A^{iA}$ & $\psi^\a_A$ & $c^a$ &
                   $\cb^a$ & $b^a$
\\ \hline
dim & $1$ & $3/2$  & $1$ & $1$ & $1$ & $3/2$ & $0$ & $2$ & $2$
\\ \hline
$\Phi\Pi$ & $0$ & $0$ & $0$ & $0$ & $0$ & $0$ & $1$ & $-1$ & $0$
\\ \hline
\end{tabular}

\vspace{.2cm}{\footnotesize {\bf Table 1.}
Dimensions and Faddeev--Popov
charges of the quantum fields.}
\end{center}

The BRS and supersymmetry operators form the algebra represented by
the relations~(2.13),~(3.3) and~(3.4). Now we collect $s$ and
$\d\equiv \eb^i Q_i$ together, defining a new operator $\QQ$
\eq
\QQ \equiv s + \eb^i Q_i + \x^\m\pa_\m - (\eb^i\g^\m\e_i) \pad{}{\x^\m}\ ,
\eqn{35}
where we introduced two global ghosts $\e_i$ and $\x^\m$, associated to
supersymmetry and translations respectively. In order to have an homogeneous
operator $\QQ$, we must assign
the following dimensions and Faddeev--Popov charges\footnote{The following
commutation rule holds for elements $\f_i$ with ghost charge $g_i$
and spinorial Grassmann parity $p_i$:
$\f_1\f_2=(-1)^{g_1g_2+p_1p_2}\f_2\f_1$.}

\begin{center}
\begin{tabular}{|l|r|r|}\hline
 & $\e_i$ & $\x^\m$
\\ \hline
dim & $-1/2$ & $-1$
\\ \hline
$\F\Pi$ & $1$ & $1$
\\ \hline
\end{tabular}

\vspace{.2cm}{\footnotesize {\bf Table 2.}
Dimensions and Faddeev--Popov
charges of the global ghosts.}
\end{center}

The quantum fields transform under $\QQ$ as follows
\eqa
\QQ A^{a} &=& f^{abc} c^b A^c + \eb^i\l^a_i + \x^\m\pa_\m A^{a} \nonumber \\
\QQ  B^{a} &=& f^{abc} c^b B^c + i\eb^i\g_5\l^a_i + \x^\m\pa_\m B^{a}
\nonumber \\
\QQ A^{a}_\m &=& -(D_\m c)^a + \eb^i\g_\m  \l^a_i + \x^\n\pa_\n A^{a}_\m
\nonumber \\
\QQ \l^a_{\a i}   &=& f^{abc} c^b \l_{\a i}^c +
           \frac{1}{2} F^a_{\m\n}(\s^{\m\n}\e_i)_\a
                  -(D_\m A)^a(\g^\m\e_i)_\a
\nonumber \\ &&                  +i(D_\m B)^a(\g^\m\g_5\e_i)_\a
                  +if^{abc}A^bB^c(\g_5\e_i)_\a + \x^\m\pa_\m \l^a_{\a i}
\nonumber \\
\QQ A^{iA}   &=& (T^a)^A_B c^a A^{iB} + \eb^i\psi^A + \x^\m\pa_\m A^{iA}
\nonumber \\
\QQ \psi^A_\a &=& (T^a)^A_B c^a \p^B_\a  -2(D_\m A^i)^A (\g^\m\e_i)_\a
                   +2(T^a)^A_B A^{iB} A^a \e_{\a i}
\\& &
                -2i(T^a)^A_B A^{iB} B^a (\g_5\e_i)_\a + \x^\m\pa_\m \psi^A_\a
\nonumber \\
\QQ c^a &=& \half f^{abc}c^bc^c
- (\eb^i\g^\m\e_{i})A^a_\m - i(\eb^i\g_5\e_{i})B^a + (\eb^i\e_{i})A^a +
\x^\m\pa_\m c^a \nonumber \\
\QQ \cb^a &=& b^a + \x^\m\pa_\m \cb^a \nonumber \\
\QQ b^a &=&  (\eb^i\g^\m\e_{i})\pa_\m\cb^a   + \x^\m\pa_\m b^a \nonumber \\
\QQ \x^\m &=& -(\eb^i\g^\m\e_{i}) \nonumber \\
\QQ \e_i &=& 0 \nonumber \ .
\eqan{36}

The action of $\QQ$ on the fields belonging to the SYM and matter multiplet
follows trivially from their BRS~(3.1) and
supersymmetry~(2.2)-(2.3)
transformations. Notice  that
we let the ghost field $c$ transform
into the field dependent gauge transformation~(2.14), in addition to
its usual BRS variation and translation.
In this way, we reach our first goal, namely the elimination
from the algebra of the field dependent gauge transformations.
The result is that the operator $\QQ$ is on--shell nilpotent
\eq
\QQ^2 = \mbox{equations of motion}\ .
\eqn{37}
Explicitely we get
\eqa
\QQ^2 \F \= 0\ \ \ \ \ \ \ \ \F = \mbox{all fields but}\ \l, \p \nonumber \\
\QQ^2 \l^a_{\a i} \= 2 \LP \eb^j \fud{S_{SYM}}{\lb^{ai}}\RP \e_{\a j}
   -\LP \eb^j \fud{S_{SYM}}{\lb^{aj}}\RP \e_{\a i} \\
\QQ^2 \p^A_\a \= -(\eb^i\e_i) \fud{S}{\pb^\a_A}
    +(\eb^i\g_5\e_i) \LP \g_5\fud{S}{\pb_A}\RP_\a
    +(\eb^i\g^\m \e_i) \LP \g_\m\fud{S}{\pb_A}\RP_\a\ .\nonumber
\eqan{38}

The operator $\QQ$ obviously describes  a symmetry of the action
$S$
\eq
\QQ S =0\ .
\eqn{39}
By means of the operator $\QQ$, which is nilpotent on all fields but the
spinors, we can construct an invariant gauge fixing term in the usual way,
namely
\eqa
S_{gf} \= \QQ \intx \cb^a \Lp \partial A^a + \frac{\t}{2} b^a \Rp \\
\= \intx \LP b^a\partial A^a +\frac{\t}{2}b^2
- (\partial^\m\cb^a)(D_\m c)^a
   +(\pa^\m\cb^a)(\eb^i\g_\m\l^a_i) \RP \ , \nonumber
\eqan{310}
where $\t$ is the gauge parameter, which is zero in the Landau gauge.

We thus end up with a symmetry of the action
\eq
S_{inv}\equiv S +S_{gf}\ ,
\eqn{311}
described by an operator, $\QQ$, which is on--shell nilpotent. This situation
is
sometimes called of Batalin -- Vilkovisky type. We can
bypass the general procedure~\cite{bv} devised to
quantize these kind of models: it is in fact  well known~\cite{qt,gms}
that, in order to write the Slavnov identity corresponding to an on--shell
nilpotent symmetry, one must add to the action terms of higher order
in the external sources which are coupled, as usual, to the nonlinear
variations of the quantum fields.

In our case the source dependent term of the action must be
\eqa
S_{ext}=\intx\LP &&\!\!\!
M^a(\QQ A^a) + N^a(\QQ B^a) + \O^{a\m}(\QQ A^a_\m)
+ \Lb^{ai}(\QQ\l^a_i) + L^a (\QQ c^a) \nonumber \\
&&\!\!\!
+ U^\ast_{iA}(\QQ A^{iA})
+ U^{iA}(\QQ A^\ast_{iA}) + \Pb_A(\QQ\p^A)
+ (\QQ\pb_A)\P^A
\nonumber\\ &&\!\!\!
- (\eb^j\L^a_i)(\Lb^{ai}\e_j) + \half (\eb^j\L^a_j)(\Lb^{ai}\e_i)
\\ &&\!\!\!
- (\eb^i \e_i) (\Pb_A \P^A) + (\eb^i\g_5\e_i) (\Pb_A\g_5\P^A)
+ (\eb^i\g^\m\e_i) (\Pb_A\g_\m\P^A) \RP\ , \nonumber
\eqan{312}
notice the nonstandard quadratic term in the external sources $\L$ and $\P$.

The total classical action
\eq
\S\equiv S_{SYM} + S_{matter} + S_{int} + S_{gf} + S_{ext}
\eqn{313}
satisfies the Slavnov identity
\eq
\SS(\S) =0\ ,
\eqn{314}
where
\eqa
\SS(\S)=\intx\LP &&\!\!\!
\fud{\S}{\O^{a\m}}
\fud{\S}{A^a_\m}
+
\fud{\S}{L^a}
\fud{\S}{c^a}
+
\fud{\S}{M^a}
\fud{\S}{A^a}
+
\fud{\S}{N^a}
\fud{\S}{B^a}
+
\fud{\S}{\Lb^{a i}}
\fud{\S}{\l^a_i}
\nonumber\\&&\!\!\!
+
\fud{\S}{U^\ast_{iA}}
\fud{\S}{A^{iA}}
+
\fud{\S}{U^{iA}}
\fud{\S}{A^\ast_{iA}}
+
\fud{\S}{\Pb_A}
\fud{\S}{\p^A}
+
\fud{\S}{\P^A}
\fud{\S}{\pb_A}
\nonumber\\&&\!\!\!
+
(b^a+\x^\m\pa_\m\cb^a)\fud{\S}{\cb^a}
+
( (\eb^i\g^\m\e_i)\pa_\m\cb^a+\x^\m\pa_\m b^a)\fud{\S}{b^a} \RP
\nonumber\\&&\!\!\!
-
(\eb^i\g^\m\e_i)\pad{\S}{\x^\m}\ .
\eqan{315}
The corresponding linearized Slavnov operator
\eqa
\BS=\intx\LP &&\!\!\!
\fud{\S}{\O^{a\m}}
\fud{}{A^a_\m}
+
\fud{\S}{A^a_\m}
\fud{}{\O^{a\m}}
+
\fud{\S}{L^a}
\fud{}{c^a}
+
\fud{\S}{c^a}
\fud{}{L^a}
+
\fud{\S}{M^a}
\fud{}{A^a}
+
\fud{\S}{A^a}
\fud{}{M^a}
\nonumber\\&&\!\!\!
+
\fud{\S}{N^a}
\fud{}{B^a}
+
\fud{\S}{B^a}
\fud{}{N^a}
+
\fud{\S}{\Lb^{a i}}
\fud{}{\l^a_i}
-
\fud{\S}{\l^a_i}
\fud{}{\Lb^{a i}}
+
\fud{\S}{U^\ast_{iA}}
\fud{}{A^{iA}}
+
\fud{\S}{A^{iA}}
\fud{}{U^\ast_{iA}}
\nonumber\\&&\!\!\!
+
\fud{\S}{U^{iA}}
\fud{}{A^\ast_{iA}}
+
\fud{\S}{A^\ast_{iA}}
\fud{}{U^{iA}}
+
\fud{\S}{\Pb_A}
\fud{}{\p^A}
-
\fud{\S}{\p^A}
\fud{}{\Pb_A}
+
\fud{\S}{\P^A}
\fud{}{\pb_A}
-
\fud{\S}{\pb_A}
\fud{}{\P^A}
\nonumber\\&&\!\!\!
+
(b^a+\x^\m\pa_\m\cb^a)\fud{}{\cb^a}
+
( (\eb^i\g^\m\e_i)\pa_\m\cb^a+\x^\m\pa_\m b^a)\fud{}{b^a} \RP
\nonumber\\&&\!\!\!
-
(\eb^i\g^\m\e_i)\pad{}{\x^\m}\
\eqan{316}
is off--shell nilpotent
\eq
\BS\BS =0\ .
\eqn{317}

The form of the linearized Slavnov operator~(3.16) implies
 that the action of $\BS$ on the quantum fields is
given by
\eq
\BS\F=\fud{\S}{K_\F}\ ,
\eqn{318}
where  $K_\F$ are the external sources coupled to the non linear
$\QQ$--variations of the quantum
fields $\F$. Therefore $\BS\F$ coincides with $\QQ\F$
for all the fields but the spinors, for which we have
\eqa
\BS \l^a_{\a i} \= \QQ\l^a_{\a i} - 2(\eb^i\L^a_i)\e_{\a j}
      +(\eb^j\L^a_j)\e_{\a i}\nonumber\\
\BS \p^A_\a \= \QQ\p^A_\a - (\eb^i \e_i) \P^A_\a
+ (\eb^i\g_5\e_i) (\g_5\P^A)_\a
+ (\eb^i\g^\m\e_i) (\g_\m\P^A)_\a\ ,
\eqan{319}
and
\eq
\BS^2\l^a_{\a i}=\BS^2\p^A_\a=0\ .
\eqn{320}
Comparing (3.20) with (3.8),
it is apparent that  the effect of adding a bilinear
term in the external sources to the action
is to modify the transformations laws of the spinor
fields in order to obtain the nilpotency. In this sense, the external sources,
whose presence is in any case necessary in view of the quantization of the
theory, in addition play the same role of the auxiliary fields usually
introduced to put the formalism off--shell.

\section{The semiclassical approximation}

The Slavnov identity~(3.14) alone is not sufficient to
uniquely determine the classical action.
The theory is defined also by the following
constraints~:
\begin{enumerate}
\item the gauge condition
\eq
\fud{\S}{b^a} = \pa A^a + \t b^a\ ;
\eqn{41}
in the Landau gauge, the commutator between the gauge condition~(4.1)
and the Slavnov identity~(3.14) gives the antighost equation
\eq
\fud{\S}{\cb^a} + \pa^\mu \fud{\S}{\O^{a\m}} -
\x^\m\pa_\m \fud{\S}{b^a} = 0\ ;
\eqn{42}
\item the $\x$--equation
\eq
\pad{\S}{\x^\m} = \D_\m\ ;
\eqn{43}
where
\eqa
\D_\m &\equiv\intx \LP & \!\!\!
- M^a\pa_\mu A^a - N^a\pa_\mu B^a - \O^{a\n}\pa_\m A^a_\n
- (\Lb^{ai}\pa_\m\l^a_i) + L^a\pa_\m c^a
\nonumber \\&&\!\!\!
 - U^\ast_{iA}\pa_\m A^{iA}
- U^{iA}\pa_\m A^\ast_{iA} - (\Pb_A\pa_\m\p^A)
+(\pa_\m\pb_A\P^A)\RP\ .
\eqan{44}
By anticommuting the $\x$--equation with the Slavnov identity~(3.14),
one gets the Ward identity for the translations
\eq
\pad{}{\x^m}\SS(\S) + \BB_\S (\pad{\S}{\x^\m} - \D_\m) =
\sum_{all\ fields\ \F}\intx (\pa_\m\F)\fud{\S}{\F} \equiv \PP_\m\S=0\ ;
\eqn{45}
\item the ghost equation, peculiar to the Landau gauge~$\t=0$~\cite{bps,gms}
\eq
\FF^a\S=\D^a\ ,
\eqn{46}
where
\eq
\FF^a\equiv\intx\LP\fud{}{c^a}+f^{abc}\cb^b\fud{}{b^c}\RP
\eqn{47}
and
\eqa
\D^a \equiv \intx\LP&&\!\!\!
f^{abc}\Lp M^bA^c + N^bB^c + \O^{b\m}A^c_\m + (\Lb^{bi}\l^c_i) - L^bc^c \Rp
\nonumber\\&&\!\!\!
-(T^a)^A_B\Lp U^\ast_{iA}A^{iB} - U^{iB}A^\ast_{iA}
+(\Pb_A\p^B) + (\pb_A\P^B)\Rp\RP\ .
\eqan{48}
The ghost equation~(4.6), anticommuted with the Slavnov
identity~(3.14),
gives the Ward identity of the rigid gauge invariance
\eq
\FF^a\SS(\S) + \BB_\S (\FF^a\S-\D^a) =
\sum_{all\ fields\ \F}\intx (\d_{rig}\F)\fud{\S}{\F} \equiv\HH^a_{rig}\S=0 \ ,
\eqn{49}
where $\d_{rig}\F$ are the rigid gauge transformations of the fields $\F$,
described by the BRS transformations~(3.1) with $c$ constant parameter.
\end{enumerate}

The constraints~(4.1), (4.3) and~(4.6)
 have the form of broken symmetries of the
action~$\S$, but notice that the breakings are linear in the quantum
fields, and therefore they are present only at the classical level,
which means that  they do not get quantum corrections.

The equations~(4.1) and~(4.3) can be solved by introducing the reduced
action $\Sh$, defined by
\eq
\S = \Sh +\intx\LP b^a\pa A^a + \frac{\t}{2}b^2 \RP + \x^\m\D_\m\ ,
\eqn{410}
where $\D_\m$ is given by~(4.4).

It follows that
\eq
\fud{\Sh}{b^a}=\pad{\Sh}{\x^\m}=0\ ,
\eqn{411}
and, for $\t$=0,
\eq
\fud{\Sh}{\cb^a} + \pa^\mu \fud{\Sh}{\O^{a\m}} =0\ ,
\eqn{412}
namely $\Sh$ does not depend on the Lagrange multiplier $b$ nor on the
global ghost $\x^\m$, and, in the Landau gauge,
it depends on the ghost $\cb$ and on the external
source $\O^\m$ only through the combination
\eq
\eta^{\a\m}\equiv\pa^\m\cb^a+\O^{a\m}\ .
\eqn{413}
The reduced action $\Sh$ is recognized to be
\eqa
\Sh \=\!\!\!\! S_{SYM} + S_{matter} + S_{int}
\nonumber \\ &&\!\!\!\!
+\intx \Lp
M^a (f^{abc} c^b A^c + \eb^i\l^a_i)
+ N^a (f^{abc} c^b B^c + i\eb^i\g_5\l^a_i)
+ \eta^{a\m} (-(D_\m c)^a + \eb^i\g_\m  \l^a_i)
\nonumber \\ &&\!\!\!\!
+ \Lb^{a\a i} (f^{abc} c^b \l_{\a i}^c +
                   \frac{1}{2} F^a_{\m\n}(\s^{\m\n}\e_i)_\a
                         -(D_\m A)^a(\g^\m\e_i)_\a
                     +i(D_\m B)^a(\g^\m\g_5\e_i)_\a
                  +if^{abc}A^bB^c(\g_5\e_i)_\a)
\nonumber \\ &&\!\!\!\!
+ L^a (\half f^{abc}c^bc^c - (\eb^i\g^\m\e_{i})A^a_\m -
i(\eb^i\g_5\e_{i})B^a +
          (\eb^i\e_{i})A^a)
\nonumber \\ &&\!\!\!\!
+ U^\ast_{iA} ( (T^a)^A_B c^a A^{iB} + \eb^i\psi^A)
+ U^{iA} ( -(T^a)^B_A c^a A^\ast_{iB} + \pb_A\e_i)
\\ &&\!\!\!\!
+ \Pb_A^\a ((T^a)^A_B c^a \p^B_\a  -2(D_\m A^i)^A (\g^\m\e_i)_\a
                   +2(T^a)^A_B A^{iB} A^a \e_{\a i}
                -2i(T^a)^A_B A^{iB} B^a (\g_5\e_i)_\a )
\nonumber \\ &&\!\!\!\!
+ (-(T^a)^B_A c^a \pb^\a_B  + 2(D_\m A^\ast_i)_A (\eb^i\g^\m)^\a
                   + 2(T^a)_A^B A^\ast_{iB} A^a \eb^{\a i}
                -2 i(T^a)_A^B A^\ast_{iB} B^a (\eb^i\g_5)^\a ) \P^A_\a
\nonumber \\ &&\!\!\!\!
- (\eb^j\L^a_i)(\Lb^{ai}\e_j) + \half (\eb^j\L^a_j)(\Lb^{ai}\e_i)
\nonumber \\ &&\!\!\!
- (\eb^i \e_i) (\Pb_A \P^A) + (\eb^i\g_5\e_i) (\Pb_A\g_5\P^A)
+ (\eb^i\g^\m\e_i) (\Pb_A\g_\m\P^A) \RP\ . \nonumber
\eqan{414}
Because of the constraints~(4.11) and~(4.12),
 the functional~$\Sh$ satisfies the
modified Slavnov identity
\eq
\SSh(\Sh) =0\ ,
\eqn{415}
where
\eqa
\SSh(\Sh)=\intx\LP &&\!\!\!
\fud{\Sh}{\eta^{a\m}}
\fud{\Sh}{A^a_\m}
+
\fud{\Sh}{L^a}
\fud{\Sh}{c^a}
+
\fud{\Sh}{M^a}
\fud{\Sh}{A^a}
+
\fud{\Sh}{N^a}
\fud{\Sh}{B^a}
+
\fud{\Sh}{\Lb^{a i}}
\fud{\Sh}{\l^a_i}
\nonumber\\&&\!\!\!
+
\fud{\Sh}{U^\ast_{iA}}
\fud{\Sh}{A^{iA}}
+
\fud{\Sh}{U^{iA}}
\fud{\Sh}{A^\ast_{iA}}
+
\fud{\Sh}{\Pb_A}
\fud{\Sh}{\p^A}
+
\fud{\Sh}{\P^A}
\fud{\Sh}{\pb_A}\RP\ ,
\eqan{416}
whose linearized Slavnov operator
\eqa
\BSh=\intx\LP &&\!\!\!\!\!
\fud{\Sh}{\eta^{a\m}}
\fud{}{A^a_\m}
+
\fud{\Sh}{A^a_\m}
\fud{}{\O^{a\m}}
+
\fud{\Sh}{L^a}
\fud{}{c^a}
+
\fud{\Sh}{c^a}
\fud{}{L^a}
+
\fud{\Sh}{M^a}
\fud{}{A^a}
+
\fud{\Sh}{A^a}
\fud{}{M^a}
\\&&\!\!\!\!\!
+
\fud{\Sh}{N^a}
\fud{}{B^a}
+
\fud{\Sh}{B^a}
\fud{}{N^a}
+
\fud{\Sh}{\Lb^{a i}}
\fud{}{\l^a_i}
-
\fud{\Sh}{\l^a_i}
\fud{}{\Lb^{a i}}
+
\fud{\Sh}{U^\ast_{iA}}
\fud{}{A^{iA}}
+
\fud{\Sh}{A^{iA}}
\fud{}{U^\ast_{iA}}
\nonumber\\&&\!\!\!\!\!
+
\fud{\Sh}{U^{iA}}
\fud{}{A^\ast_{iA}}
+
\fud{\Sh}{A^\ast_{iA}}
\fud{}{U^{iA}}
+
\fud{\Sh}{\Pb_A}
\fud{}{\p^A}
-
\fud{\Sh}{\p^A}
\fud{}{\Pb_A}
+
\fud{\Sh}{\P^A}
\fud{}{\pb_A}
-
\fud{\Sh}{\pb_A}
\fud{}{\P^A}\RP\nonumber
\eqan{417}
is such that
\eq
\BSh\Sh=0
\eqn{418}
and therefore it is nilpotent
\eq
\BSh\BSh=0\ .
\eqn{419}

Finally, in this framework the ghost equation of the Landau gauge becomes
\eq
\FF^a\longrightarrow\hat\FF^a=\intx\fud{}{c^a}\ .
\eqn{420}

\section{Conclusions}

In this paper we gave an off--shell formulation of $N=2$ Super Yang -- Mills
theories, without using auxiliary fields. The main result we obtained
is represented by the Slavnov identity~(3.14)
\eq
\SS(\S)=0\ ,
\eqn{51}
which is a generalized version
of the ordinary one corresponding to the BRS symmetry. It  describes
the invariance of the theory not only under the BRS
transformations~(3.1),
but also under the supersymmetry~(2.2)-(2.3) and the
translations~(4.5). From~(5.1) follows the off--shell nilpotency
of the linearized Slavnov operator
\eq
\BS\BS=0\ ,
\eqn{52}
which contains all the
informations concerning the algebraic structure of the theory,
namely the nilpotency of the BRS operator and the {\it off--shell} closure
of the supersymmetry commutators on the translations.
This result has been obtained
without making use of auxiliary fields and therefore this method is quite
general, being applicable also when no off--shell formulation in terms
of auxiliary fields exists, like the $N=4$ case. The point is that the problem
of finding auxiliary fields is circumvented in this formulation, because
in order to study the quantum extension
of a theory characterized by nonlinear symmetries, it is necessary to couple
external sources to the nonlinear variations of the quantum fields. These
external sources transform into the equations of motion of the corresponding
quantum fields, therefore in this framework they do precisely the same
job as the conventional auxiliary fields. Even when a complete set of auxiliary
fields can be introduced, as it happens for the $N=1$ and $N=2$ cases,
the alternative formulation described in this paper avoids the inconvenients
arising from the infinite chain of external sources introduced to put the
formalism off--shell, thus making easy the otherwise impossible quantization
of the model.

Moreover, for what concerns the renormalization of the theory,
the lack of an acceptable regularization preserving both BRS and
supersymmetry renders the algebraic method of renormalization even more
necessary than usual~\cite{bm}. Formally, the $N=2$ SYM
theory is now completely described by a Slavnov identity and the constraints
discussed in Section~4. The algebraic renormalization of such a model, although
technically rather involved, can now be carried on as if we were dealing with
an ordinary gauge field theory~\cite{wp}.

\vspace{2cm}
{\bf Acknowledgments}
I would like to thank  A.~Blasi, O.~Piguet and M.~Porrati
 for useful discussions, and D.~Balboni for his hints on
spinorial algebra.

\end{document}